\documentclass[twocolumn,amsmath,amssymb,showpacs]{revtex4}
\usepackage{subfigure}
\usepackage{wrapfig}
\usepackage{color}
\usepackage{graphicx}
\usepackage{bm}
\usepackage{amsmath}
\usepackage{amsfonts}
\usepackage{amssymb}
\makeatletter
 
\begin{document}
\title{Roles of Dry Friction in Fluctuating Motion of Adiabatic Piston}
\author{Tomohiko G. Sano and Hisao Hayakawa}
\affiliation{Yukawa Institute for Theoretical Physics, Kyoto University, Kitashirakawa-oiwake cho, Sakyo-ku, Kyoto 606-8502, Japan}

\begin{abstract}
The motion of an adiabatic piston under dry friction is investigated to clarify the roles of dry friction in non-equilibrium steady states. We clarify that dry friction can reverse the direction of the piston motion and causes a discontinuity or a cusp-like singularity for velocity distribution functions of the piston. We also show that the heat fluctuation relation is modified under dry friction.
\end{abstract}
\maketitle

\section{Introduction}
Recent developments in experimental technique enable us to control small systems and non-equilibrium systems, such as nano-scale systems, single colloidal systems and biological systems, to clarify their thermodynamic structures, in detail \cite{optical, textbook, small_sys}.
One of the most important applications of manipulation techniques for small systems is the design of nano-machines or sub-micron machines \cite{nano_motor, nano_motor_rev1,nano_motor_rev2}. The difficulty to realize efficient small machines is the existence of dry friction when two solids are in contact, because the dry friction wears down the small machines \cite{nano_friction}. Thus, to control systems under dry friction is indispensable to make small machines. Furthermore, dry friction cannot be ignored in the study of non-equilibrium steady states, such as transport phenomena of molecular motor and motor proteins, because there are many unavoidable obstacles which play central roles in small realistic systems, such as dry friction, wear, adhesion, electrification, and so on \cite{motor_electro,granick_1,granick_2, friction_text}. Indeed, dry friction is known to play a central role in the directed motion of Kinesin motors on micro-tubes \cite{protein, protein_1}. Experiments for macroscopic systems under dry friction reveal that the dry friction has an important role to rectify unbiased fluctuations, i.e. to extract work from an equilibrium environment \cite{andrea_prl, andrea_2013, andrea_rapid}. The motor interacting with its supporting axis via dry friction rotates even in an equilibrium fluid. Recent studies on Brownian motion under dry friction clarifies that the particle motion is characterized by non-Gaussian statistics \cite{degennes, k_hayakawa, hayakawa,menzel_g,solich1,solich2, talbot1,talbot2, touchette_fp, andrea_epl, pathint}.

Although dry friction plays essential roles in non-equilibrium transport \cite{nano_friction,protein, protein_1,friction_text,granick_1,granick_2} and it is ubiquitous throughout nature from a biological surface to an atomic-scale surface \cite{bio_interface,prl_2013,prl_2011}, the energetics for the systems under dry friction has been elusive so far. For systems without dry friction, there exists the energetics in the Langevin description so called stochastic energetics \cite{s_e1, s_e2, s_e3}, in which the first law of thermodynamics holds at the level of single fluctuating particle manipulation. The original form of stochastic energetics has been restricted to systems of a fluctuating single particle driven by Gaussian white noise, while it is extended to those driven by non-Gaussian white noise by introducing the new stochastic products \cite{ksh_1, ksh_2}. 

In this paper, we study energy transfer, such as momentum or heat transfer, for systems with a fluctuating boundary under dry friction. For this purpose, we study the motion of an adiabatic piston under the mechanical equilibrium, which is located between two equilibrium environments characterized by two different temperatures and densities. Lieb suggested that the equilibrium thermodynamics cannot tell us whether the adiabatic piston moves or not \cite{adp0,adp1}. This problem is solved analytically by using Boltzmann-Lorentz equation \cite{gp} and is recently phenomenologically understood through the concept of the momentum transfer deficit due to dissipation (MDD) \cite{sekimoto}. However, the motion of the adiabatic piston under dry friction is little known. 

Let us clarify the difference between this paper and previous studies \cite{talbot2, solich1,solich2}. Although the roles of dry friction in the asymmetric granular piston with the different restitution coefficient have already been discussed in Ref. \cite{talbot2}, its roles in the symmetric piston exposed to two thermal gases of different temperatures have not been analyzed yet. Baule and Sollich have studied a solvable model for a fluctuating piston whose two faces are respectively kicked by a single state-independent Poissonian noise under dry friction, assuming an exponential distribution for the amplitude and the constant event probability for each noise \cite{solich1,solich2}. However, the motion of the piston surrounded by the two thermal gases, which are characterized by state-dependent compound Poissonian noises, under dry friction has not been analyzed yet.

\begin{figure}[h]
\begin{center}
\includegraphics[scale = 0.4]{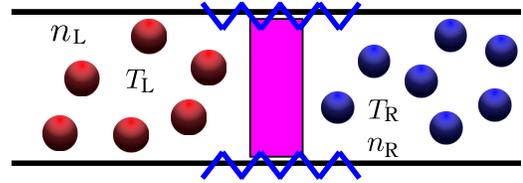}
\caption{(Color online) Schematic picture for the system with a fluctuating boundary under dry friction. Blue zigzag lines represent dry friction. Ideal gas molecules are enclosed in a container and the piston with a finite mass $M$ separate gas into two regions. Gas densities $n_{\rm L}, n_{\rm R}$ and temperatures $T_{\rm L}, T_{\rm R}$ are assumed to be constants.}
\label{setup}
\end{center}
\end{figure}

The analysis of the fluctuating motion of adiabatic piston is important on the construction of engines for realistic small systems. Indeed, small heat engines should also include the fluctuating motion of an adiabatic piston, to separate the system from external environment, in a similar manner to the macroscopic engines \cite{broeck}. As the first step to consider the energetics for realistic systems, we study the motion of an adiabatic piston under dry friction.

The organization of this paper is as follows. Firstly, we introduce our setup and the basic stochastic equation of motion, in which the piston is kicked by double Poissonian noises from left and right sides (see Fig. \ref{setup}) in Sec. \ref{s_setup}. We prove that the introduced equation is equivalent to the Boltzmann-Lorentz equation. In Sec. \ref{s_result}, we show main results on the velocity distribution function, the reverse motion of the piston and the fluctuation relation for the work done by gas under dry friction. Our theoretical results are verified through the numerical calculation of the stochastic equation of motion. Lastly, we conclude the paper in Sec. \ref{s_concl}. In Appendix A, the detailed derivation of the fluctuation relation is presented.

\section{Setup}\label{s_setup}
Let us enclose ideal gas molecules of mass $m$ in a container and put a piston of mass $M$ separating the gas into two parts. In Fig. \ref{setup}, the densities and the temperatures of separated gases in left and right sides are represented by $n_{\rm L}, T_{\rm L}$ and $n_{\rm R}, T_{\rm R}$, respectively. Here, we assume that the temperatures and the densities near the piston are unaffected by the existence of the fluctuating boundary. We also assume that the piston moves only in the horizontal direction (see Fig. 1) under the influence of dry dynamical friction. Moreover, molecules are assumed to be in equilibrium, while the collisions between molecules and the piston are characterized by the restitution coefficient $e$, because the piston is composed of a collection of molecules. We introduce the stochastic equation of motion for the piston as follows:
\begin{equation}
M\frac{d\hat{V}}{dt} = \hat{F}_{\rm L} + \hat{F}_{\rm R} + \hat{F}_{\rm fri}, \label{eom}
\end{equation}
where $\hat{F}_{\alpha} (\alpha = {\rm L\ or\ R})$ is the stochastic force acting on the piston due to the kick from $\alpha$ side of the piston, and $\hat{V}$ denote the stochastic velocity of the piston. We assume that the stochastic forces $\hat{F}_{\alpha}$ can be described by the state-dependent compound white Poissonian process:
\begin{eqnarray}
\hat{F}_{\alpha} &\equiv& \sum_v P_v (\hat{V}) \cdot \hat{\xi}_{\alpha} ^v (t|\hat{V}) \label{model1},\\
P_v (\hat{V}) &\equiv& \frac{1+e}{2}\frac{2\epsilon^2}{1+\epsilon^2} M(v - \hat{V}) \label{model3},
\end{eqnarray}
$(\alpha = {\rm L\ or\ R})$, where $\hat{\xi}_{\alpha} ^v (t|\hat{V})$ is one-sided Poissonian noise whose probability is equivalent to collision probability for gas molecules of the velocity between $v$ and $v + dv$ on the piston: 
\begin{equation}
\lambda_{v} ^{\alpha} \equiv dv |v - \hat{V}| \cdot \Theta(\varepsilon^{\alpha}(\hat{V}-v)) {n}_{\alpha}A \phi(v, {T}_{\alpha}).
\end{equation}
Here, we have introduced the area of the piston $A$, Maxwell distribution $\phi(v,T) \equiv \sqrt{m/2\pi k_B T}\exp(-mv^2 /2k_B T)$, Boltzmann constant $k_B$, and Heaviside function $\Theta(x) = 1 (x \geq 0)$ and $\Theta(x) = 0 (x < 0)$ with $\varepsilon^{\rm L} \equiv -1$\ {\rm and}\ $\varepsilon^{\rm R} \equiv +1$. $P_v$ represents the one-dimensional momentum change of the piston for each collision between the gas molecule of velocity $v$ and the piston. The symbol ``\ $\cdot$\ " in Eq. (\ref{model1}) represents It\^{o} product \cite{gardiner,kampen}. We assume that $\epsilon \equiv \sqrt{m/M}$, the mass ratio between molecules and the piston, is small but finite.
Here, the piston is assumed to move along the container under the influence of dry friction from the side walls 
\begin{equation}
\hat{F}_{\rm fri} \equiv - \epsilon \bar{F}_{\rm fri} \sigma(\hat{V}), \label{frihat}
\end{equation}
where $\sigma(x) = x/|x|$ is the sign function \cite{degennes, k_hayakawa, hayakawa,menzel_g,solich1,solich2, talbot1,talbot2}, and $\bar{F}_{\rm fri}$ will be determined later. 
For later discussion, we assume that the mechanical balance condition between two gases is always satisfied: $P \equiv n_{\rm L} T_{\rm L} = n_{\rm R} T_{\rm R}$. 

To examine our theoretical consideration below, we adopt the velocity Verlet method for time integration of  Eq. (\ref{eom}) with time interval $dt/t_0 = 0.01$, where we have introduced $t_0 \equiv x_0/v_{T_{\rm R}}$, and $x_0 \equiv Mv_{T_{\rm R}}^2 /PA$. We discretize the jump rates $\lambda_{v} ^{\rm \alpha}$ by replacing $dv$ by $\Delta v_{\alpha} = v_{T_{\alpha}}/50$ and $v$ by $v_i$ with $-10 v_{T_{\alpha}} < v_i <10 v_{T_{\alpha}}$ for $\alpha = {\rm L\ or\ R}\ {\rm and}\ 1 \leq i \leq 1000$, with the thermal velocity $v_T \equiv \sqrt{2k_B T/M}$ of the temperature $T$. $e = 0.9$ and $\epsilon = 0.1$ are fixed for our simulations.

The time evolution of VDF for the piston driven by Eqs. (\ref{model1}) and (\ref{model3}) under the dry friction (\ref{frihat}) satisfies the Boltzmann-Lorentz equation \cite{gardiner, kampen,brilli}:
\begin{eqnarray}
\frac{\partial f(V,t)}{\partial t} + \frac{\partial}{\partial V} \left\{\frac{F_{\rm fri}}{M}f(V,t)\right\}&=& J_{\rm L} + J_{\rm R},\label{mb}
\end{eqnarray}
as shown in Ref. \cite{kampen} and in the next paragraph. We have introduced the collision integral $J_{\alpha} (\alpha = {\rm L\ or\ R})$ as
\begin{widetext}
\begin{equation}
J_{\alpha} \equiv n_{\alpha} A \int dv |v - V| \{ \Lambda  \Theta(\varepsilon^{\alpha}(V''- v '')) f(V'',t)\phi(v '', T_{\alpha}) - \Theta(\varepsilon^{\alpha}(V-v))f(V,t)\phi(v, T_{\alpha}) \},
\end{equation}
\end{widetext}
where $v''$ and $V''$ represent the pre-collision velocities of the molecule vertical to the piston and those of the piston, respectively, which lead to the corresponding velocities $v$ and $V$, and $\Lambda \equiv 1/e^2$.

Let us prove the equivalency between the stochastic equation of motion Eqs. (\ref{eom}) - (\ref{frihat}) and the Boltzmann-Lorentz equation (\ref{mb}). For an arbitrary analytic function $h = h(\hat{V})$, its differentiation $dh(\hat{V}) \equiv h(\hat{V} + d\hat{V}) - h(\hat{V})$ can be represented as
\begin{widetext}
\begin{eqnarray}
dh(\hat{V}) &=& \sum_{n=1} ^{\infty} \frac{(d\hat{V})^n}{n!}\cdot \left. \frac{\partial^n h}{\partial V^n} \right|_{V = \hat{V}}\nonumber\\
&=& \sum_{n=1} ^{\infty} \frac{1}{n!}{\left\{\sum_v \left(\frac{P_v}{M} \cdot d\hat{L}_{\rm L} ^v\right) + \sum_v\left(\frac{P_v}{M} \cdot d\hat{L}_{\rm R} ^v\right) + \frac{\hat{F}_{\rm fri}}{M} dt \right\}^n}\cdot \left. \frac{\partial^n h}{\partial V^n} \right|_{V = \hat{V}}\nonumber\\
&=& \sum_{n = 1} ^{\infty} \frac{1}{n!} \sum_{\alpha = {\rm L, R}}\left\{ \sum_v \left(\frac{P_v}{M}\cdot d\hat{L}_{\alpha} ^v\right)^n\right\} \cdot \left. \frac{\partial^n h}{\partial V^n} \right|_{V = \hat{V}} + dt \frac{\hat{F}_{\rm fri}}{M}\cdot  \left. \frac{\partial h}{\partial V} \right|_{V = \hat{V}} + o(dt)\label{dh},
\end{eqnarray}
\end{widetext}
where we substitute $d\hat{V} = \sum_v (P_v /M) \cdot (d\hat{L}_{\rm L} ^v + d\hat{L}_{\rm R} ^v) + \hat{F}_{\rm fri} dt/M$ into the Taylor expansion of $h$ and pick up only $O(dt)$ terms. Here, we have introduced the total differentiation $d\hat{L}_{\alpha} ^v$ of $L_{\alpha} ^v (t|\hat{V}) \equiv \int_0 ^t \hat{\xi}_{\alpha} ^v(s|\hat{V})ds\ (\alpha = {\rm L\ or\ R})$, noting that $(d\hat{L}_{\alpha} ^v)^n = O(dt)$, $d\hat{L}_{\rm L} ^v\cdot d\hat{L}_{\rm R} ^v = o(dt)$ and $d\hat{L}_{\alpha} ^v\cdot d\hat{L}_{\alpha} ^{v'} = o(dt)$ for $v \ne v'$. The ensemble average and the partial integral of Eq. (\ref{dh}) leads to
\begin{widetext}
\begin{eqnarray}
\frac{\partial f}{\partial t} + \frac{\partial}{\partial V} \left\{\frac{F_{\rm fri}}{M}f(V,t)\right\}&=& \sum_{n = 1} ^{\infty} \frac{(-1)^n}{n!}\frac{\partial ^n}{\partial V^n} \left\{\sum_{\alpha = {\rm L, R}} \sum_v\left(\frac{P_v}{M}\right)^n \lambda_{v} ^{\alpha} f\right\}\nonumber\\
&=& \sum_{n=1} ^{\infty} \frac{(-1)^n}{n!}\frac{\partial^n}{\partial V^n} \left[\left\{n_{\rm L}A\int_V ^{\infty} dv\left(\frac{P_v}{M}\right)^n |v - V| \phi(v, T_{\rm L})\right\}f \right. \nonumber\\ && \left.+ \left\{n_{\rm R}A\int_{-\infty} ^V dv\left(\frac{P_v}{M}\right)^n |v - V| \phi(v, T_{\rm R})\right\}f \right],\label{fp_inf}
\end{eqnarray}
\end{widetext}
where we have used the martingale property of It\^o product as $\langle ( {P_v(\hat{V})}/{M})^n \cdot d\hat{L}_{\alpha} ^v \rangle = \langle({P_v (\hat{V})}/{M})^n \rangle\langle (d\hat{L}_{\alpha} ^v)^n\rangle = ({P_v ({V})}/{M})^n \lambda_v ^{\alpha}dt$.
The last equation in Eq. (\ref{fp_inf}) is well known to be derived through Kramers-Moyal expansion of the right hand side of Eq. (\ref{mb}) \cite{kampen}. Thus, Eqs. (\ref{eom}) - (\ref{frihat}) are equivalent to Boltzmann-Lorentz equation Eq. (\ref{mb}). 

\section{Main Results}\label{s_result}
In this section, we present main results on the velocity distribution function of the piston in \ref{s_result1}, the steady state velocity in \ref{s_result2}, and, the fluctuation relation for the work done by the gas in \ref{s_result3}.
\subsection{Velocity Distribution Function}\label{s_result1}
Expanding Eq. (\ref{mb}) or (\ref{fp_inf}) in terms of a small but a finite parameter $\epsilon$ \cite{brilli}, we obtain Fokker-Planck-like equation for $f = f(V,t)$ up to $O(\epsilon^2)$:
\begin{eqnarray}
\frac{\partial f}{\partial t} &=& \epsilon \frac{\gamma_0}{M}\left[ \frac{\partial }{\partial V} \left\{V + \mu_0  v_{T_e}\sigma(V) \right\}f + \frac{ v_{T_e}^2}{2} \frac{\partial^2f}{\partial V^2} \right]\nonumber\\ &&+ \epsilon^2 {C}\frac{\gamma_0}{M}\left[ \frac{\partial}{\partial V}\frac{V^2}{v_{T_e}}f
- \frac{v_{T_e}^3}{4}\frac{\partial ^3 f }{\partial V^3}\right] + O(\epsilon^3) \label{fok_p}.
\end{eqnarray}
Here, the first two terms on the right hand side of Eq. (\ref{fok_p}) proportional to the first derivative term in Eq. (\ref{fok_p}) represents the force. Thus, the proportional constant of the friction force in Eq. (\ref{frihat}) can be determined as
\begin{equation}
\bar{F}_{\rm fri} \equiv \mu_0 \gamma_0 v_{T_e}
\end{equation}
with $\gamma_0 \equiv \gamma_L + \gamma_R$,
\begin{eqnarray}
\gamma_{\alpha} &\equiv& \frac{2(1+e)}{\sqrt{{\pi}}} \frac{PA}{v_{T_{\alpha}}}
\end{eqnarray}
$(\alpha = {\rm L\ or\ R})$, the effective temperature $T_e \equiv (1+e)\sqrt{T_{\rm L} T_{\rm R}}/2$ and the friction constant $\mu_0$.  The steady state VDF $f_{\rm SS}(V)$ up to $O(\epsilon)$ can be readily obtained from Eq. (\ref{fok_p}):
\begin{eqnarray}
f_{\rm SS}(V) &=& {(1 + \epsilon a_1(V) + O(\epsilon^2))}f_0 (V),\label{st_vdf_fri}\\
f_0 (V) &\equiv& \frac{1}{Z} \exp\left[-\frac{M}{2k_B T_e} (V^2 + 2\mu_0v_{T_e} |V|)\right],\\
a_1(V) &\equiv& C \left\{ -{\mu_0} \sigma(V) \left(\frac{MV^2}{k_B T_e} -1 \right) \right. \label{a1}\\ && \left. + \left(1-2{\mu_0 ^2}\right)\frac{V}{v_{T_e}} - \frac{V^3}{3v_{T_e} ^3}\right\}, \nonumber\\
C &\equiv& \sqrt{\pi T_e} \left(\frac{1}{\sqrt{T_{\rm L}}} - \frac{1}{\sqrt{T_{\rm R}}}\right),\label{c_def}
\end{eqnarray}
where we have introduced the normalized constant $Z \equiv  \sqrt{\pi}v_{T_e} e^{\mu_0 ^2} {\rm erfc}(\mu_0)$, It should be noted that the restitution coefficient only appears through $\gamma_0$ and $T_e$. 

\begin{figure}[h]
\begin{center}
\includegraphics[scale = 0.7]{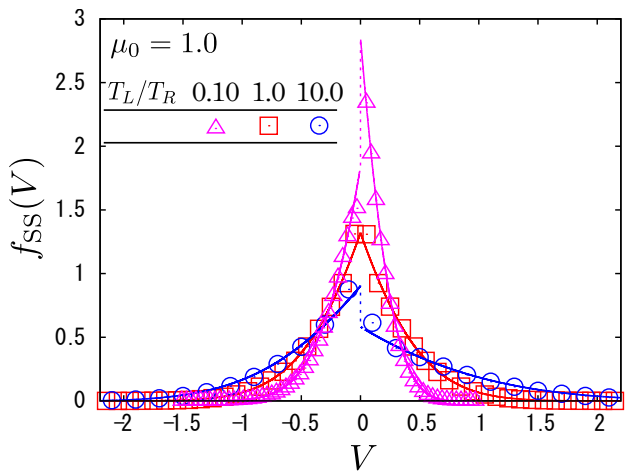}
\caption{(Color online) The obtained steady state VDFs Eqs. (\ref{st_vdf_fri}) - (\ref{c_def}) for $\mu_0 = 1.0$ and $e = 0.9$ are verified through the simulation of Eq. (\ref{eom}). We average the data over 1000 ensembles with the time average for $0 < t/t_0 < 400$. Purple triangles, red squares and blue circles are data for $T_{\rm L}/T_{\rm R} = 0.10, 1.0, 10.0$, respectively, where the corresponding theoretical curves are represented by solid lines and dashed lines denote discontinuity at $V = 0$.}
\label{vdf_fig}
\end{center}
\end{figure}

References \cite{talbot1,talbot2,solich1,solich2} reports the existence of the discontinuity and the cusp singularity in VDFs of a stochastic motion of the piston under dry friction. As we expected, we obtain the consistent results with those in the previous studies, i.e. there exists a discontinuity at $V = 0$ for $T_{\rm L} \ne T_{\rm R}$, and the cusp-like singularity appears at $V = 0$ for $T_{\rm L} = T_{\rm R}$. The obtained singularity is close to that in Ref. \cite{talbot2, solich1}, while the singularities appear at $V \ne 0$, in addition to $V=0$ in Ref. \cite{solich2}. We note that the amount of gap at $V = 0$ increases linearly with $\mu_0$.

We numerically solve Eq. (\ref{eom}) for $0 < t/t_0 < 400$ and average the data over 1000 ensembles, to obtain the data for VDF and compare it with Eq. (\ref{st_vdf_fri}) in Fig. \ref{vdf_fig} for $e = 0.9$ and $\mu_0 = 1.0$. As can be seen from Fig. \ref{vdf_fig}, it is obvious that our theory precisely reproduces the results of the simulation.

\begin{figure}[h]
\begin{center}
\includegraphics[scale = 0.7]{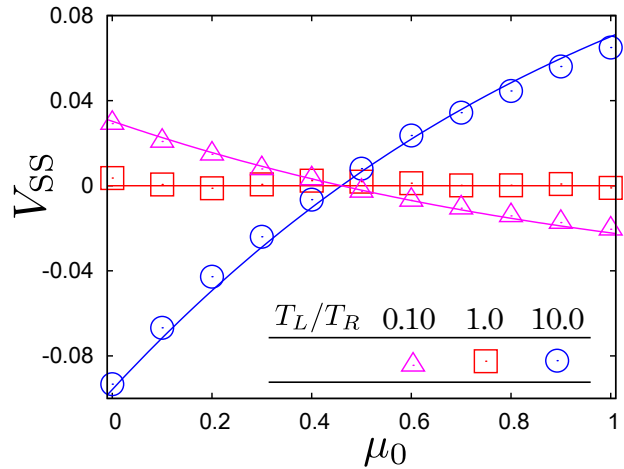}
\caption{(Color online) Reverse motion of the adiabatic piston against the friction constant $\mu_0$ is verified for $e = 0.9$. We numerically solve Eq. (\ref{eom}) and take steady state average for $0 < t/t_0 < 400$. The numerical data are obtained from the ensemble average over 1000 samples. Purple triangles, red squares and blue circles are data for $T_{\rm L}/T_{\rm R} = 0.10, 1.0, 10.0$, respectively, where the corresponding theoretical curves are represented by solid lines.}
\label{reverse}
\end{center}
\end{figure}

\subsection{Reverse motion of Adiabatic Piston}\label{s_result2}
It is known that the piston moves toward the high temperature side under the condition $n_{\rm L} T_{\rm L} = n_{\rm R} T_{\rm R}$ and $T_{\rm L} \ne T_{\rm R}$. As will be shown, however, the direction of the piston motion can be reversed under the dry friction. Indeed, the averaged steady state velocity of the piston defined by ${V}_{\rm SS}  \equiv \int dV Vf_{\rm SS}(V)$ is given by:
\begin{eqnarray}
{V}_{\rm SS} &=& V_{\rm ad} \left\{1 + 4\mu_0 \left(\mu_0 - \frac{\mu_0 ^3}{3} - \frac{7}{6\sqrt{\pi}e^{\mu_0 ^2}{\rm erfc}(\mu_0)} \right. \right. \nonumber\\ && \left. \left. +  \frac{\mu_0 ^2}{3\sqrt{\pi}e^{\mu_0 ^2}{\rm erfc}(\mu_0)}\right) \right\} +O(\epsilon^2) \label{rev},
\end{eqnarray}
where we have introduced $V_{\rm ad}$ as the steady velocity of the piston without any dry friction:
\begin{equation}
V_{\rm ad} \equiv \epsilon \frac{{\sqrt{\pi}}}{{4}} v_{T_e} ^2 \left(\frac{1}{v_{T_{\rm L}}} - \frac{1}{v_{T_{\rm R}}}\right).
\end{equation}
The notable fact in Eq. (\ref{rev}) is that the direction of the piston motion is changed around $\mu_0 \simeq 0.46$ (see Fig. \ref{reverse}).

The validity for Eq. (\ref{rev}) is verified through the direct simulation of Eqs. (\ref{eom}) - (\ref{frihat}) in Fig. \ref{reverse}, where we average the data for $0 < t/t_0 < 400$ and the ensemble average is taken over 1000 samples. As can be seen in Fig. \ref{reverse}, Eq. (\ref{rev}) reproduces the accurate behavior of Eq. (\ref{eom}). 

Through the expansion in terms of $\epsilon$ up to $O(\epsilon^2)$, Eq. (\ref{eom}) under the steady state average is reduced to
\begin{eqnarray}
0 &=& - \epsilon \gamma_0 V_{\rm SS} + \epsilon^2 \frac{C\gamma_0}{2v_{T_e}} \left\langle V^2 \right\rangle_{\rm SS} - \epsilon^2 C \bar{F}_{\rm fri} \check{f}(\mu_0) \label{mdd1},
\end{eqnarray}
where we have introduced the positive function
\begin{equation}
\check{f}(\mu_0) \equiv \frac{\mu_0}{2} + \frac{\mu_0 ^3}{3} + \frac{2 - \mu_0 ^2}{3\sqrt{\pi} {\rm erfc}(\mu_0) e^{\mu_0 ^2}} > 0.
\end{equation}
Here, the second term on the right hand side of Eq. (\ref{mdd1}) is the force due to MDD \cite{sekimoto}, and the direction of the steady friction force is opposite to MDD, from which the change of direction of the piston motion originates. As the friction force becomes larger, it can be shown that the sign of $V_{\rm SS}$ is switched, because $\left\langle V^2 \right\rangle_{\rm SS} > 0$ and $\check{f}(\mu_0) > 0$. Thus, in contrast to systems without any dry friction, the direction of the piston motion under the dry friction does not correspond to that of the force due to MDD.
\subsection{Fluctuation Relation under dry friction}\label{s_result3}
\begin{figure*}
\begin{center}
\includegraphics[scale = 1.2]{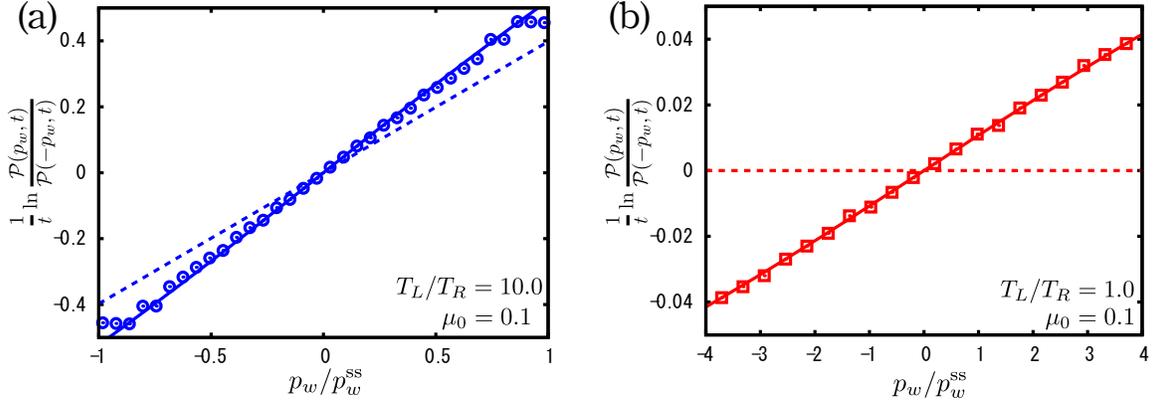}
\caption{(Color online) The fluctuation relation under dry friction Eq. (\ref{ft_fri}) is verified through our simulation for $\mu_0 = 0.1, e = 0.9$ and $t/t_0 = 20.0$, where theoretical curves are represented by solid lines. The number of samples is $2.5 \times 10^5$. Blue circles in (a) and red squares in (b) are the numerical data for $T_{\rm L}/T_{\rm R} = 10.0, 1.0$, respectively. The blue and red dashed lines are theoretical lines without dry friction for $T_{\rm L}/T_{\rm R} = 10.0, 1.0$, respectively. The events satisfying $p_w < - p_w ^{\rm ss}$ for $T_{\rm L}/T_{\rm R} = 10.0$ are so rare events that they could not be detected through our calculation, while numerical data for $T_{\rm L}/T_{\rm R} = 1.0$ reproduces the theoretical curve for large $|p_w / p_w ^{\rm ss}|$}
\label{ft_fri_fig}
\end{center}
\end{figure*}
Let us discuss the large deviation property \cite{touchette} for the work done by the system under dry friction. Fluctuation relation is one of the universal relations in non-equilibrium systems found in the last a few decades \cite{evans, gc,kurchan,lebowitz_s,visco}. The fluctuation relation for frictionless granular systems has been reported recently \cite{HO}. 
The fluctuation relations under dry friction are derived for the work done by the non-fluctuating external system under the dry friction in Ref. \cite{touchette_nonlinear}, and experimentally discussed in Ref. \cite{andrea_2013, lohse, langmuir}.
However, the work done by the fluctuating gas under dry friction has not been investigated. Here, we derive a fluctuation relation for the work done by the gas under dry friction, considering the excess work defined by $d\hat{W} ' _{\rm L} \equiv d\hat{W}_{\rm L} - F_0 \hat{V}dt$, ${d\hat{W}_{\rm L}}/dt \equiv \sum_v \{M(V^2 - V''^{2})/2\} \cdot \hat{\xi}_{\rm L} ^v (t|\hat{V})$, with the pre-collisional velocity $V''$ and $F_0 \equiv \langle \hat{F}_{\rm L} (V = 0) \rangle = (1+e)PA/\{2(1+\epsilon^2)\}$ in terms of the perturbation of small $\mu_0$, as shown in Appendix. \ref{s_derive}. 
Introducing the distribution for the excess power ${\mathcal P}(p_{w}, t) \equiv \langle\delta(\hat{W}_{\rm L}'(t) - p_{w}t)\rangle$, we obtain the fluctuation relation under dry friction up to $O(\epsilon, \mu_0)$:
\begin{eqnarray}
\lim_{t \to \infty} \frac{1}{t} \ln \frac{{\mathcal P}(p_{w}, t)}{{\mathcal P}(-p_{w},t)} &=& \Delta \beta_e p_{w} + \frac{\bar{F}_{\rm fri} }{PA}B(p_{w}) p_{w} \nonumber \\ && + O(\epsilon^2, \mu_0 ^2),\label{ft_fri}
\end{eqnarray}
where we have introduced the difference of inverse temperatures
\begin{eqnarray}
\Delta \beta_e &\equiv& \frac{2}{1+e} \left(\frac{1}{k_B T_{\rm R}} - \frac{1}{k_B T_{\rm L}}\right),
\end{eqnarray}
and the nonlinear function of $p_w$
\begin{eqnarray}
B(p_{w}) &\equiv& \sqrt{\frac{v_{T_{\rm L}}}{v_{T_e}}} \frac{\sqrt{\eta^*(p_{w})}}{(1+e)k_B T_e}\\
\eta^{*2}(p_{w}) &\equiv& \frac{1 + (\tilde{T} ^{1/4} - \tilde{T} ^{-1/4})^2 }{1 + 2{\pi} ({p_{w}}/{\epsilon PA v_{T_{e}}} )^2/(1+e)^3}
\end{eqnarray}
with $\tilde{T} \equiv T_{\rm L}/T_{\rm R}$. See Appendix A for the derivation of the fluctuation relation. We solve Eq. (\ref{eom}) with $\mu_0 = 0.1$ and $e = 0.9$ for $0 < t/t_0 \leq 20.0$ to verify the validity of Eq. (\ref{ft_fri}) as shown in Fig. \ref{ft_fri_fig} (a) for $T_{\rm L}/T_{\rm R} = 10.0$ and (b) for $T_{\rm L}/T_{\rm R} = 1.0$, where the number of samples is $2.5 \times 10^5$.
Blue circles of Fig. \ref{ft_fri_fig} (a) and red squares of Fig. \ref{ft_fri_fig} (b) are the numerical data for $T_{\rm L}/T_{\rm R} = 10.0, 1.0$, respectively, and the solid curves denote the corresponding theoretical curves. The blue and red dashed lines are theoretical lines without dry friction for $T_{\rm L}/T_{\rm R} = 10.0, 1.0$, respectively. Here we use the scaled $p_w$ by the corresponding steady state value $p_w ^{\rm ss} = 0.2099$ for $T_{\rm L}/T_{\rm R} = 10.0$ and $p_w ^{\rm ss} = 0.01281$ for $T_{\rm L}/T_{\rm R} = 1.0$, respectively. We only plot the data for $|p_w/p_w ^{\rm ss}| < 1$ for $T_{\rm L}/T_{\rm R} = 10.0$ because the events satisfying $p_w < - p_w ^{\rm ss}$ are so rare events that they could not be detected through our calculation. On the other hand, numerical data for $T_{\rm L}/T_{\rm R} = 1.0$ reproduce the theoretical curve even for large $|p_w / p_w ^{\rm ss}|$.

\section{Concluding Remarks}\label{s_concl}
In this paper, we clarified the role of dry friction in the fluctuating motion of an adiabatic piston surrounded by two thermal temperatures. Through the analysis of the Boltzmann-Lorentz equation Eq. (\ref{mb}), we found the singularities only at $V = 0$ as those in Ref. \cite{talbot2, solich1}, while they are different from those in Ref. \cite{solich2}. VDF of a fluctuating piston has a cusp-like singularity for $T_{\rm L} = T_{\rm R}$ and a discontinuity at $V = 0$ for $T_{\rm L} \ne T_{\rm R}$, as in Eqs. (\ref{st_vdf_fri}) - (\ref{c_def}) and Fig. \ref{vdf_fig}. We obtained the friction dependence of the velocity of the piston motion in Eq. (\ref{rev}), whose direction is changed above the threshold of the friction const $\mu_0$, as in Fig. \ref{reverse}. The change of the direction of the piston motion has not been reported in the previous studies for the fluctuating piston under dry friction \cite{talbot2,solich1,solich2}. We also demonstrated that the conventional fluctuation relation for the fluctuating work is modified due to the existence of dry friction.

\begin{acknowledgements}
We would like to thank K. Kanazawa and M. Itami for fruitful discussions.
This work is partially supported by the Grant-in-Aid of MEXT (Grants No. 25287098) and for the Global COE program ``The Next Generation of Physics, Spun from Universality and Emergenceh from MEXT, Japan.
\end{acknowledgements}

\appendix
\section{Derivation of Fluctuation Relation under Dry Friction}\label{s_derive}
In this appendix, we derive the fluctuation relation under dry friction Eq. (\ref{ft_fri}), writing $\hat{W} \equiv \hat{W}_{\rm L}'$ and $d\hat{W} \equiv d\hat{W}_{\rm L} - F_0 \hat{V} dt$. Let us derive a Master equation for $f(V,W,t) \equiv \langle \hat{f}(\hat{V}(t), \hat{W}_{\rm L} '(t))\rangle$ with $\hat{f}(\hat{V}(t), \hat{W}_{\rm L} '(t)) \equiv \delta(V - \hat{V}(t)) \delta(W - \hat{W}_{\rm L} '(t))$, following Ref. \cite{gardiner}. For an arbitrary function $g = g(\hat{V}, \hat{W})$, the differentiation $dg(\hat{V}, \hat{W}) \equiv g(\hat{V} + d\hat{V}, \hat{W} + d\hat{W}) - g(\hat{V}, \hat{W})$ 
 is given by
 \begin{widetext}
 \begin{eqnarray}
dg(\hat{V}, \hat{W}) &=& \frac{1}{M} \left(\sum_v d\hat{L}_{\rm L} ^v \cdot P_v\right) \cdot \left\{\left(\frac{\partial}{\partial V} + MV \frac{\partial}{\partial W} \right)g \right\} + \frac{1}{2M^2}  \left(\sum_v d\hat{L}_{\rm L} ^v \cdot P_v ^2 \right) \cdot \left\{\left(\frac{\partial}{\partial V} + MV \frac{\partial}{\partial W} \right)^2g \right\} \nonumber\\
&& +  \frac{1}{M} \left(\sum_v d\hat{L}_{\rm R} ^v \cdot P_v\right) \cdot \frac{\partial g}{\partial V}+ \frac{1}{2M^2}  \left(\sum_v d\hat{L}_{\rm R} ^v \cdot P_v ^2 \right) \cdot \frac{\partial^2 g}{\partial V^2} - F_0 \hat{V}\cdot \frac{\partial g}{\partial W} dt - \epsilon \frac{\bar{F}_{\rm fri}}{M} \sigma(V)\cdot \frac{\partial g}{\partial V} dt \nonumber\\ && + O(\epsilon^2 dt) \label{diff}
\end{eqnarray}
\end{widetext}
It should be noted that $\langle \sum_v d\hat{L}_{\alpha} ^v\cdot P_v ^{n+1} \rangle = O(\epsilon^n dt)$. By taking the ensemble average of Eq. (\ref{diff})  and expanding it up to $O(\epsilon)$, the Master equation for $f(V,W,t)$ is derived \cite{gardiner}.
Introducing Laplace transformation of $f(V,W,t)$ as
\begin{equation}
\tilde{f}_{\beta} \equiv \int dW e^{-\beta W} f(V,W,t),
\end{equation}
we obtain the time evolution for $\tilde{f}_{\beta}$:
\begin{eqnarray}
\frac{\partial}{\partial t} \tilde{f}_{\beta} = \epsilon \frac{\gamma_0}{M} \left(L_{\beta} + L_{\rm fri} \right)\tilde{f}_{\beta} + O(\epsilon^2), \label{lap_fp}
\end{eqnarray}
where $L_{\beta}$ and $L_{\rm fri}$ denote the linear operators on $\tilde{f}_{\beta}$ as
\begin{widetext}
\begin{eqnarray}
L_{\beta} &=& \frac{v_{T_e} ^2}{2}\frac{\partial^2}{\partial V^2} + \left(1 + \check{\beta} \frac{\gamma_{\rm L}}{\gamma_0} \sqrt{\frac{T_{\rm L}}{T_{\rm R}}}\right) \frac{\partial}{\partial V} V -\frac{\check{\beta}}{2} \frac{\gamma_{\rm L}}{\gamma_0} \sqrt{\frac{T_{\rm L}}{T_{\rm R}}} +  \frac{\gamma_{\rm L}}{\gamma_0} \check{\beta}\left(1 + \sqrt{\frac{T_{\rm L}}{T_{\rm R}}}\frac{\check{\beta}}{4}\right) \frac{V^2}{v_{T_e} ^2}\label{fp_1st},\\
L_{\rm fri} &\equiv& \mu_0 v_{T_e}\frac{\partial}{\partial V} \sigma(V)
\end{eqnarray}
\end{widetext}
with $\check{\beta} \equiv 2k_B T_e \beta$. The eigenvalues $\kappa_n$ and eigenfunctions for the operator (\ref{fp_1st}) and its adjoint operator $L^{\dagger} _{\beta}$ are discussed in Ref. \cite{visco}
:
\begin{eqnarray}
L _{\beta} \psi_n (V) &=& \kappa_n \psi_n (V)\\
L_{\beta} ^{\dagger}{\phi}_n (V) &=& \kappa_n \phi_n(V)
\end{eqnarray}
\begin{eqnarray}
\psi_n (V) &=& \sqrt{\frac{\zeta}{{2\pi v_{T_e}{2^n n!}}}} {\exp\left[-\frac{\zeta V^2}{2v_{T_e} ^2 }\right]}{H_n \left( \frac{\sqrt{\eta}V}{v_{T_e}} \right)},\\
 \phi_n (V) &=& \sqrt{\frac{2\eta}{\zeta v_{T_e}}} \exp\left[ -\left(\eta -\frac{\zeta}{2} \right) \frac{V^2}{v_{T_e} ^2} \right] H_n \left( \frac{\sqrt{\eta}V}{v_{T_e}} \right), \nonumber\\
\end{eqnarray}
\begin{eqnarray}
\kappa_n(\beta) &=& \frac{1}{2} \left\{1 - (1+2n) \eta(\beta) \right\},\\
\eta(\beta) &\equiv& \sqrt{1 + \frac{\sqrt{\tilde{T}}(2k_B T_e)^2}{(1 + \sqrt{\tilde{T}})^2} \beta(\Delta \beta_e - \beta),}
\end{eqnarray}
where Hermite polynomials are defined as $H_n (x) \equiv (-1)^n e^{x^2} (d/dx)^n e^{-x^2}$ and $\int_{-\infty} ^{\infty} dx e^{-x^2}H_n(x) H_l (x) = \sqrt{\pi}2^n n! \delta_{nl} (n, l = 0, 1, \cdots)$.
$\kappa_n(\beta)$ has Gallavotti-Cohen-type symmetry as $\kappa_n \left(\Delta \beta_e - \beta \right) = \kappa_n(\beta)$ \cite{gc, lebowitz_s}, which leads to the conventional fluctuation relation without dry friction:
\begin{eqnarray}
\lim_{t \to \infty} \frac{1}{t} \ln \frac{{\mathcal P}(p_w, t)}{{\mathcal P}(-p_w,t)} = \Delta \beta_e p_w + O(\epsilon^2).
\end{eqnarray}
Let us solve eigenvalue problem for $L_{\beta} + L_{\rm fri}$ perturbatively up to $O(\epsilon, \mu_0)$, assuming that $\mu_0$ is small:
\begin{eqnarray}
\left({L}_{\beta} + L_{\rm fri}\right) \bar{\psi}_n (V) &=& \bar{\kappa}_n(\beta) \bar{\psi}_n (V),\label{1p}
\end{eqnarray}
We assume that ${\rm Re}(\bar{\kappa}_n) \leq {\rm Re}(\bar{\kappa}_m)$ for $n > m$, where ${\rm Re}(a)$ represents the real part of any complex number $a$. Multiplying $\phi_n(V)$ on both sides of Eq. (\ref{1p}), integrating them over $V$ and substituting $\bar{\kappa}_n(\beta) = \kappa_n(\beta) + \mu_0 \kappa_n ^{(1)}(\beta) + O(\mu_0^2)$, $\bar{\psi}_n (V) = {\psi}_n (V) + O(\mu_0)$ into Eq. (\ref{1p}) for $n = 0$, we obtain:
\begin{eqnarray}
\mu_0 \kappa_0 ^{(1)}(\beta)
&=& - \frac{\mu_0}{\sqrt{\pi \eta(\beta)}} \left(1 + \check{\beta} \frac{\sqrt{\tilde{T}}}{1+\sqrt{\tilde{T}}} \right).
\end{eqnarray}
The largest eigenvalue of the operator $\epsilon \gamma_0 (L_{\beta} + L_{\rm fri})/M$ is known to be equal to the scaled cumulant generating function \cite{nemoto}:
\begin{equation}
\lim_{t \to \infty} \frac{1}{t}\ln \langle e^{-\beta \hat{W}_{\rm L} '(t)} \rangle = \epsilon \frac{\gamma_0}{M}\bar{\kappa}_0(\beta)
\end{equation}
Thus, according to Ref. \cite{touchette}, the large deviation property for $\hat{W}_{\rm L} '$ under the dry friction is characterized by the Legendre transformation of the maximum eingenvalue of ${L}_{\beta} + L_{\rm fri}$:
\begin{equation}
\lim_{t \to \infty} \frac{1}{t} \ln {\mathcal P}(p_w,t) = p_w \beta^* + \epsilon \frac{\gamma_0}{M} \bar{\kappa}_0(\beta^*), 
\end{equation}
where $\beta^* = \beta^*(p_w)$ gives the minimum for $p_w \beta + \epsilon {\gamma_0} \bar{\kappa}_0(\beta)/M$. Taking the asymmetric part in terms of $p_w$, we obtain Eq. (\ref{ft_fri}).


\end{document}